\documentclass[preprint,showpacs,aps,a4paper]{revtex4}
\usepackage{graphicx}
\usepackage{amssymb,amsmath}
\begin{document}

\title{Projected single-spin flip dynamics  in the Ising Model}

\author{A. L. C. Ferreira }
\affiliation{Departmento de  F\'\i sica, Universidade de Aveiro,
 3810-193 Aveiro, Portugal}

\author{ Ra\'ul Toral}
\affiliation{Instituto Mediterr\'aneo de
Estudios Avanzados (IMEDEA) UIB-CSIC,  Ed. Mateu Orfila, Campus UIB, E-07122 Palma de Mallorca, Spain}

\date{\today}

\begin{abstract}
We study transition matrices for projected dynamics  in the
energy-magnetization space, magnetization space and energy space.
Several single spin flip dynamics are considered such as the
Glauber and Metropolis canonical ensemble dynamics and the
Metropolis dynamics for three multicanonical ensembles: the flat
energy-magnetization histogram, the flat energy histogram and the
flat magnetization histogram. From the numerical diagonalization
of the matrices for the projected dynamics we obtain  the
sub-dominant eigenvalue and the largest relaxation times  for
systems of varying size. Although, the projected dynamics is an
approximation to the full state space dynamics comparison with
some available results, obtained by other authors, shows that
projection in the magnetization space is a reasonably accurate
method to study the scaling of relaxation times with system size.
The transition matrices for arbitrary single-spin flip dynamics
are obtained from a single Monte-Carlo estimate of the infinite
temperature transition-matrix, for each system size, which makes
the method an efficient tool to evaluate the relative performance
of any arbitrary local spin-flip dynamics. We also present new
results for appropriately defined average tunnelling times of
magnetization and compute their finite-size scaling exponents that
we compare with results of energy tunnelling exponents available
for the flat energy histogram multicanonical ensemble.

\end{abstract}

\maketitle

\section{Introduction}
The dynamical critical behavior of statistical physics
models is a problem that
attracts considerable attention\cite{HH77,Sokal97,Nightingale96,Nightingale00}. From a fundamental point of view
one is interested in the identification and characterization of
the different dynamical universality classes, known to be more
restricted than the static ones. Different algorithms for
canonical ensemble simulations have been proposed belonging to
different universality classes\cite{cluster,Wolff89}. Still, increasing
relaxation times with system size are a major limitation to the
statistical precision of the numerical estimates obtained in the
simulations. New algorithms, aiming to estimate the number of
states of a given energy, have also been
proposed\cite{multicanonical,lee93,bhanot}. These algorithms simulate a
multicanonical ensemble with the advantage that a single simulation provides
information on the properties of the system in a wide temperature
range. However, such algorithms also suffer from slowing down with
increasing system size and the study of their dynamical properties with simple
and efficient methods is essential to ascertain their relative
performance.

Many numerical methods have been used to study stochastic dynamics
of statistical physics models. These methods measure the largest
relaxation time of the dynamics, a time which increases with
system size according to dynamic finite-size scaling theory. The
exact diagonalization of the transition matrix in the full state
space can be done only for very small systems. To overcome this
limitation, one can instead estimate by Monte-Carlo methods the
auto-correlation function of the slowest observable in the system,
that whose long time behavior gives the largest relaxation time.
Although this method is free of systematic errors, one needs to
consider very long simulation runs to get a reasonably small
statistical error in the auto-correlation function. Several other
methods have been used, including a variational
technique\cite{Nightingale96,Nightingale00}  allowing the
estimation of the sub-dominant eigenvalue of the full state-space
transition matrix.

{\sl Projected dynamics} was proposed to study metastability and
nucleation in the Ising
model\cite{Lee95,Shteto97,Shteto99,Kolesik97,Kolesik98,Novotny01}.
The idea behind this method is to derive a dynamics in a
restricted space of one or several variables. Choosing
appropriately such variables and neglecting non-Markovian memory
terms one hopes that the resulting approximated Markovian dynamics
is a good approximation to the full state space dynamics. The
usefulness of the method has been proved in the context of the
study of metastability in the Ising model where the direct dynamic
Monte-Carlo simulation is unable to bare with the large time-scale
of the problem\cite{Novotny01}. The non-lumpability of the the
full state-space transition rate matrix
with respect to energy and magnetization classification of the
states leads to the upcoming of memory terms when projecting the
dynamics in these restricted spaces\cite{Kemeny76, Novotny01}. To recover the Markovian
character of the dynamics, these memory terms are neglected and the
resulting projected dynamics becomes only approximated.

In this article we study the projected dynamics  behavior for the
square lattice, nearest-neighbor, Ising model, in the energy and
magnetization spaces for
two local spin flip algorithms. Namely, the Glauber and
the Metropolis et al.\cite{MetGl} critical canonical ensemble dynamics, and
three multicanonical algorithms: the flat energy-magnetization
histogram,  the flat energy histogram and the flat
magnetization histogram dynamics. Although the dynamics associated with the transition rate
matrices in these restricted spaces are only approximate, we show, by
comparison with full state space results, that they can be used to
get reasonably accurate estimates of the dynamical properties.
From the numerical diagonalization of these matrices, and the determination of their sub-dominant
eigenvalue, we compute the largest relaxation times for
systems of varying size. The method proposed can be applied to
other models and other dynamics thus leading to a simple and
efficient estimation of the scaling with system size of the
largest relaxation time. Such studies are needed to assess the
relative performance of Monte-Carlo simulation algorithms.

Projected dynamics transition rate matrices were also considered
in the context of the transition matrix
Monte-Carlo\cite{Wang99,Wang01}. Using an acceptance
probability written in terms of the infinite temperature energy
space transition matrix it is possible to perform simulations that
visit with equal probability the spectra of energies of the model
thus doing flat energy histogram simulations. For the case of the
Ising model that we consider in this work this algorithm is
easily generalized to  simulations with a flat energy and
magnetization histogram. We use this flat energy-magnetization
histogram ensemble to numerically estimate the infinite
temperature transition rate matrix in the space of energy and
magnetization from which all the results presented in this work
were derived.

For  multicanonical algorithms, average tunnelling times between
the ground-state and states with higher energy (for example zero
energy) have been considered\cite{dayal2004}. It has been shown
that these tunnelling times may scale differently with system size
when we consider going up  (from a low energy to a high energy) or
going down in the energy\cite{costa2005}. We present new results
for average tunnelling times of magnetization, in several
multicanonical ensembles, using projected dynamics, that show a
similar behavior and that can be compared with results of other
authors for tunnelling times in the energy space.

The new method proposed by us, to study approximately the local
dynamics, is efficient because: (1) the dynamic exponents
estimates are reasonably accurate when compared with corresponding
quantities obtained by other methods; (2) any, arbitrary,
single-spin flip dynamics can be studied from a single Monte-Carlo
estimation of an infinite temperature transition matrix in the
energy-magnetization space (corresponding to acceptance of all the
proposed configurations); the consideration of a specific dynamics
comes only from the weighting of this matrix with the
corresponding acceptance probability; (3)  the dimensional
reduction achieved by the projection allows the application of
matrix diagonalization techniques for bigger system sizes.

The outline of the paper is as follows: In section \ref{sectionII}
we discuss the projection procedure, in section \ref{sectionIII}
we show how the infinite temperature transition matrix is computed
from Monte-Carlo simulations for different system sizes and we
define the projected transition matrices for the different
ensembles and dynamics considered, in section \ref{sectionIV} we
present results for the largest relaxation times and the
corresponding dynamical exponents, in section \ref{sectionV} we
define and compute tunnelling times in the magnetization space and
their finite-size scaling exponents and, finally, in section
\ref{sectionVI} we summarize our main conclusions.

\section{Projected dynamics}
\label{sectionII}
 The Markov chain master equation in the full
state space is:

\begin{equation}
\frac{dP(\vec \sigma,t)}{dt}=\sum_{\vec \sigma'} [W(\vec \sigma,\vec \sigma') P(\vec \sigma',t)
- P(\vec \sigma,t) W(\vec \sigma',\vec \sigma)],
\label{eq:1}
\end{equation}
where $ \vec \sigma$ denotes a state of the system, $P(\vec
\sigma,t)$ is the probability for the system to be in a given state
at time $t$ and $W(\vec \sigma,\vec \sigma')$ is the transition
rate from state $\vec \sigma'$  to $\vec \sigma$. In the case of
an Ising model $ \vec \sigma\equiv (\sigma_1,
...,\sigma_N)$ specifies the state of each of  $N$ spins of the
system, $\sigma_i$, that can take two values, $\sigma_i=\pm 1$. The transition
rate obeys detailed balance

\begin{equation}
P_{st}(\vec \sigma) W(\vec \sigma',\vec \sigma)=P_{st}(\vec
\sigma') W(\vec \sigma,\vec \sigma') \label{eq:2}
\end{equation}
relatively to a stationary
distribution, $P_{st}(\vec \sigma)$, which we consider to be an arbitrary function $P_{st}(E(\vec \sigma),M(\vec \sigma))$, of the energy $E(\vec
\sigma)=-\sum_{\langle i,j\rangle} \sigma_i \sigma_j$ (where the sum is over
all neighbor pairs, $\langle i,j\rangle$), and the magnetization $M(\vec
\sigma)=\sum_i \sigma_i$.

The detailed balance equation can be summed up relative to all
$\vec \sigma$  states with a given energy $E=E(\vec \sigma)$ and magnetization $M=M(\vec \sigma)$, and all $\vec \sigma'$ states with energy
$E'=E(\vec \sigma')$ and magnetization $M'=M(\vec \sigma')$, to obtain:

\begin{eqnarray}
\sum_{\vec \sigma,\vec \sigma'} P_{st}(\vec \sigma) W(\vec
\sigma',\vec \sigma) \delta_{E,E(\vec \sigma)} \delta_{E',E(\vec
\sigma')} \delta_{M,M(\vec \sigma)} \delta_{M',M(\vec \sigma')} &=&\\
\nonumber \sum_{\vec \sigma,\vec \sigma'} P_{st}(\vec \sigma')
W(\vec \sigma,\vec \sigma') \delta_{E,E(\vec \sigma)}
&\delta_{E',E(\vec \sigma')} &\delta_{M,M(\vec \sigma)}
\delta_{M',M(\vec \sigma')} \label{eq:3}
\end{eqnarray}
being $\delta_{a,b}$ the Kronecker delta. Since the stationary distribution is assumed to be a function of the energy  and magnetization only, it can be
taken out of the summation, giving

\begin{equation}
p(E,M) T(E',M';E,M)= p(E',M') T(E,M;E',M')
\label{eq:4}
\end{equation}
where $p(E,M)=P_{st}(E,M) \Omega(E,M)$ is the stationary probability for the
system to have energy $E$ and magnetization $M$, and $\Omega(E,M)$
is the number of states with  energy $E$ and magnetization $M$. In this expression, we have defined,

\begin{equation}
T(E',M';E,M)=\frac{1}{\Omega(E,M)}
\sum_{\vec \sigma,\vec \sigma' }
W(\vec \sigma',\vec \sigma)
\delta_{E,E(\vec \sigma)} \delta_{E',E(\vec \sigma')} \delta_{M,M(\vec \sigma)} \delta_{M',M(\vec \sigma')}
\label{eq:5}
\end{equation}
as the transition matrix between energy and magnetization states $(E,M)$
and $(E',M')$.

Summing up the master equation in the same way we would obtain the evolution equation for the time dependent probability $p(E,M,t)$ for the system to have energy $E$ and magnetization $M$ at time $t$:
\begin{equation}
\frac{dp(E,M,t)}{dt}=\sum_{E',M'} [T(E,M;E',M';t) p(E',M',t)
- p(E,M,t) T(E',M';E,M;t)],
\label{eq:6}
\end{equation}
with a time-dependent transition matrix:
\begin{equation}
T(E',M';E,M;t)=\frac{1}{p(E,M,t)}
\sum_{\vec \sigma,\vec \sigma'}
P(\vec \sigma,t) W(\vec \sigma',\vec \sigma)
\delta_{E,E(\vec \sigma)} \delta_{E',E(\vec \sigma')} \delta_{M,M(\vec \sigma)} \delta_{M',M(\vec \sigma')}
\label{eq:7}
\end{equation}
This time dependent matrix approaches the transition rate matrix in Eq. (\ref{eq:5}) for
large times when $P(\vec \sigma,t)/p(E,M,t) \rightarrow 1/\Omega(E,M)$.
The so-called {\sl projected dynamics}  neglects this time dependence and
considers instead the Markov process associated with $T(E',M';E,M)$:
\begin{equation}
\frac{dp(E,M,t)}{dt}=\sum_{E',M'} [T(E,M;E',M') p(E',M',t)
- p(E,M,t) T(E',M';E,M)].
\label{eq:6a}
\end{equation}
Starting with the projection operator technique, in a discrete
time formulation, the approximation can be regarded as equivalent
to dropping out some memory terms\cite{Shteto99}. Note that the
dynamics of the Markovian process associated with these transition
matrices would be equivalent to the full state space dynamics if
it were lumpable\cite{Kemeny76} with respect to a classification
of the states in terms of energy and magnetization. However, this
is known not to be the case for canonical ensemble
dynamics\cite{Novotny01}, although the flat magnetization ensemble
that we study later is lumpable with respect to a magnetization
classification of the states.

Further projection on the energy space can be done by summing for
all $M$ and $M'$ the detailed balance condition in the $E,M$ space
(Eq.\ref{eq:4}) :
\begin{equation}
p(E)
\sum_{M,M'}\frac{p(E,M)}{p(E)}T(E',M';E,M)=p(E')\sum_{M,M'}\frac{p(E',M')}{p(E')},
T(E,M;E',M') \label{eq:7a}
\end{equation}
which is a detailed balance relation $p(E)T(E',E)=p(E')T(E,E')$ in the energy space with a
projected transition matrix
\begin{equation}
T(E';E)= \sum_{M,M'}\frac{p(E,M)}{p(E)}T(E',M';E,M). \label{eq:7b}
\end{equation}

Note that for the ensembles where $P_{st}(\vec \sigma)$
depends just on the energy (and not on the magnetization) the previous expression can be simplified
to:

\begin{equation}
T(E';E)=\frac{1}{\Omega(E)} \sum_{M,M'} \Omega(E,M) T(E',M';E,M)=
\frac{1}{\Omega(E)} \sum_{\vec \sigma,\vec \sigma' } W(\vec
\sigma',\vec \sigma) \delta_{E,E(\vec \sigma)} \delta_{E',E(\vec
\sigma')} \label{eq:8}
\end{equation}
with $\Omega(E)=\sum_M\Omega(E,M)$ is the number of states with energy $E$. If $P_{st}(\vec \sigma)$ depends on energy and magnetization
simultaneously the above simplification can not be done.

The transition matrix $T(E,E')$ can be used to define a Markov chain dynamics in the restricted energy space:
\begin{equation}
\frac{dp(E,t)}{dt}=\sum_{E'} [T(E;E') p(E',t)
- p(E,t) T(E';E)].
\label{eq:6b}
\end{equation}

In the same way we can obtain a detailed balance relation in the
magnetization space:
\begin{equation}
p(M) \sum_{E,E'}\frac{p(E,M)}{p(M)}T(E',M';E,M)=p(M')\sum_{E,E'}\frac{p(E',M')}{p(M')}
T(E,M;E',M'),
\label{eq:9}
\end{equation}
which is a detailed balance relation $p(M)T(M',M)=p(M')T(M,M')$ in the magnetization space with a
projected transition matrix
\begin{equation}
T(M';M)= \sum_{E,E'}\frac{p(E,M)}{p(M)}T(E',M';E,M).
\label{eq:10}
\end{equation}
The transition matrix $T(M,M')$ can be used to define a Markov chain dynamics in the restricted magnetization space:
\begin{equation}
\frac{dp(M,t)}{dt}=\sum_{M'} [T(M;M') p(M',t)
- p(M,t) T(M';M)].
\label{eq:6c}
\end{equation}
Nevertheless the approximation assumed in the projected dynamics,
the detailed balance relations satisfied by the transition matrices
defined above assure that the long time behavior of the related
stochastic processeses defined by Eqs. (\ref{eq:6a}),  (\ref{eq:6b}) and (\ref{eq:6c}) are still characterized by the correct stationary
probability distributions $p(E,M)$, $p(E)$ and $p(M)$, respectively.

In the following sections, we study single spin flip dynamics in
the canonical ensemble characterized by the stationary
distribution at inverse temperature $\beta$, $P_{st}(\vec
\sigma)=\exp(-\beta E(\vec \sigma))/Z$ as well as three
multicanonical ensembles with flat energy-magnetization, flat
energy and flat magnetization histograms with $P_{st}(\vec
\sigma)=1/\Omega(E,M)$, $P_{st}(\vec \sigma)=1/\Omega(E)$ and
$P_{st}(\vec \sigma)=1/\Omega(M)$, respectively. Note that
$\Omega(M)=\sum_E\Omega(E,M)$ is exactly known to be
$\Omega(M)=\binom{N}{\frac{N+M}{2}}$ and that  an efficient
numerical scheme (not used by us in the present work) developed by
Beale \cite{Beale96} allows to compute exactly $\Omega(E)$ for the
two-dimensional Ising model for moderate system sizes $N$.

\section{Numerical calculation of transition matrices}
\label{sectionIII}
We now explain our method to compute numerically the transition matrices $T(E',M';E,M)$, $T(E',E)$ and $T(M',M)$ defined in Eqs. (\ref{eq:5}), (\ref{eq:7b}) and (\ref{eq:10}), respectively. We start by recalling that for single spin flip dynamics the transition rate $W(\vec \sigma',\vec \sigma)$ can be separated in a proposal
step and an acceptance step. In the proposal step we choose, with
equal probability, one of the spins of the system and propose to
flip it. Thus a given system state may have a non-zero transition
rate to $N$ other system states that differ in the state of a single
spin. In the acceptance step we accept the proposed configuration
with a probability $a(E',M';E,M)$ that we assume depends only on the energy and
magnetization of the initial and final configurations.

Consider the detailed balance relation (\ref{eq:4}) when we accept all the proposed configurations.
This is the case, for example, in the Metropolis {\sl et al.} algorithm at infinite temperature.
The probability to measure an energy $E$ and magnetization $M$ is
then equal to $\Omega(E,M)/2^N$ since all states have equal
probability. Thus we can write the relation,
\begin{equation}
\Omega(E,M) T_\infty(E',M';E,M)= \Omega(E',M') T_\infty(E,M;E',M')
\label{eq:11}
\end{equation}
known as the broad histogram equation\cite{bhm}. For a general
single spin flip algorithm characterized by $a(E',M';E,M)$we can
write,
\begin{equation}
T(E',M';E,M)=  T_\infty(E',M';E,M) a(E',M';E,M).
 \label{eq:12}
\end{equation}

The numerical determination of $T_\infty(E',M';E,M)$ can be done from the estimator:
\begin{equation}
T_\infty(E',M';E,M)= \frac{1}{N H_{sim}(E,M)} \sum_{k=1}^{N_m}
N(\vec \sigma_k,\Delta E, \Delta M) \delta_{E,E(\vec \sigma_k)}
\delta_{M,M(\vec \sigma_k)} \label{eq:13}
\end{equation}
where the summation is done over the $N_m$ configurations
generated by the Monte-Carlo procedure and $N(\vec \sigma_k,\Delta
E,\Delta M)$ is the number of configurations with energy
$E'=E+\Delta E$ and magnetization $M'=M+\Delta M$ that can be
obtained from configuration $\vec \sigma_k$ by flipping a single
spin and the quantity $H_{sim}(E,M)$ is the energy and
magnetization histogram of the simulation.  The estimator in Eq.
(\ref{eq:13}) is related to the average of $\langle N(\vec \sigma,\Delta
E,\Delta M)\rangle_{E,M}$ in the constant energy and magnetization
ensemble,
\begin{eqnarray}
\langle N(\vec \sigma, \Delta E, \Delta M)\rangle_{E,M}&=&\frac{1}{\Omega(E,M)}
\sum_{\vec \sigma} N(\vec \sigma,\Delta E,\Delta M)
\delta_{E,E(\vec \sigma)} \delta_{M,M(\vec \sigma)}
\\ \nonumber
&=&\langle N(\vec \sigma,\Delta E,\Delta M) \delta_{E,E(\vec \sigma)}
\delta_{M,M(\vec \sigma)}/ (P_{sim}(\vec \sigma)
\Omega(E,M))\rangle_{sim} \label{eq:14}
\end{eqnarray}
where $P_{sim}(\vec \sigma)$ is the probability to visit a
particular state in the simulation ensemble whose averages are
denoted by, $\langle \dots\rangle_{sim}$. Since $H_{sim}(E,M)= N_m P_{sim}(\vec
\sigma) \Omega(E,M)$ with $P_{sim}(\vec \sigma)$ dependent only on
E and M, we see that Eq. (\ref{eq:13}) provides the correct
estimator. For the two-dimensional square lattice,
nearest-neighbor, Ising model each spin can have between zero and
and four nearest neighbors in the same state of the spin. When
this spin flips there are five possible energy changes, $\Delta E$
and two magnetization changes, $\Delta M$. Thus, one needs to
count the number of spin flips that lead to a energy and
magnetization change in each of these possible ten classes.

In this work we have estimated $ T_\infty(E',M';E,M)$ by doing transition matrix
Monte-Carlo
simulations in a two-dimensional, nearest neighbor, Ising model of size $N=L^2$ with an acceptance probability given by,
$a(E',M';E,M)=\min\left(1,\frac{T_\infty(E,M;E',M')}{T_\infty(E',M';E,M)}\right)$.
From eqs. (\ref{eq:4}) and (\ref{eq:12}) we can see that this
choice leads to a flat energy and magnetization histogram. The
algorithm starts with an initial estimate of
$T_\infty(E',M';E,M)$ that is improved as more configurations are
generated. We have used the n-fold way simulation algorithm of
Kalos and Lebowitz\cite{kalos,Wang01} and the number of simulated spin flips per
number of spins was $10^8$ for each of the systems studied, $L=3,
...,21,30$. Note that, when one
considers a n-fold way simulation, the
histogram of energy and magnetization, $H_{sim}(E,M)$ is the
average time spent in a given value of energy and magnetization
that may differ from the average number of hits to a particular
energy and magnetization value. In this case, the expression
(\ref{eq:13}) should be modified  to weight each of the generated
configurations  with the estimated average time spent in these
configurations (a small but systematic error arises in the
results if this weighting is not done).

The projected transition matrices in the energy-magnetization
space, $T(E',M';E,M)$ are obtained from the simulation estimates
of $T_\infty(E',M';E,M)$ by using eq. (\ref{eq:12}). We consider
the following dynamics: (1) the Metropolis canonical ensemble
dynamics with  $a(E',M';E,M)=\min\left(1,\exp(-\beta
(E'-E))\right)$ (2) the Glauber canonical ensemble dynamics with
$a(E',M';E,M)=\frac{1}{2}\left(1-\tanh(\frac{\beta}{2}(E'-E))\right)$
(3) the flat energy and magnetization histogram Metropolis
dynamics with
$a(E',M';E,M)=\min\left(1,\frac{T_\infty(E,M;E',M')}{T_\infty(E',M';E,M)}\right)$,
(4) the Metropolis flat energy dynamics also known as entropic
sampling with
$a(E',M';E,M)=\min\left(1,\frac{\Omega(E)}{\Omega(E')}\right)$ and
(5) the Metropolis flat magnetization dynamics with
$a(E',M';E,M)=\min\left(1,\frac{\Omega(M)}{\Omega(M')}\right)$.

For the energy magnetization-space with dimension $(N+1)^2 \times
(N+1)^2$ we have obtained results from the diagonalization of
$T(E',M';E,M)$ up to $N=8^2$. For all the system sizes studied we
have found the stationary probablities $p(E,M)$ after solving
numerically for the steady state regime of the system of equations
Eq.(\ref{eq:6a})\footnote{The solution has been found by an
iterative method in which, at  each iteration the values of the
energy-magnetization probability at $(E_1,M_1)$ and $(E_1,-M_1)$
were kept constant with a value $p(E_1,M_1)=p(E_1,-M_1)=1/2$. This
procedure  was found to improve considerably the convergence. The
values of $(E_1,M_1)$ were chosen to be near the $(E,M)$ region
where $p(E,M)$ has an appreciable value. The iteration was stopped
when the measured relative change of $\sum_{E,M}p(E,M)$ did not
change by more than $10^{-12}$.} :

\begin{equation}
\sum_{E',M'} [T(E,M;E',M') p(E',M')
- p(E,M) T(E',M';E,M)]=0.
\label{eq:14a}
\end{equation}

For the flat energy histogram dynamics we
need to know $\Omega(E)$  to construct the corresponding
acceptance probability. This quantity can be obtained from
$\Omega(E,M)$ after the solution of the homogeneous linear system
of equations
\begin{equation}
\sum_{E',M'} [T_\infty(E,M;E',M') \Omega(E',M')
- \Omega(E,M) T_\infty(E',M';E,M)]=0.
\label{eq:14b}
\end{equation}

Note that it is possible to compute,
\begin{equation}
T_\infty(E';E)=\sum_{M',M}  \frac{\Omega(E,M)}{\Omega(E)}
T_\infty(E',M';E,M), \label{eq:14c}
\end{equation}
and write
$a(E',M';E,M)=\min\left(1,\frac{T_\infty(E;E')}{T_\infty(E';E)}\right)$ for a
flat energy histogram ensemble which is completely equivalent to
$a(E',M';E,M)=\min\left(1,\frac{\Omega(E)}{\Omega(E')}\right)$.

\section{Largest Relaxation times}
\label{sectionIV}
 We have considered  a discrete time transition
matrix defined as $\gamma(E,M;E'M')=T(E,M;E'M')$ for $(E,M)\ne(E',M')$ and
$\gamma(E,M;E,M)=1-\sum_{E',M'}T(E',M';E,M)$ for  $(E,M)\equiv
(E',M')$.   This corresponds to the Markov chain equation
$p(E,M,t+1)=\sum_{E',M'}\gamma(E,M;E'M') p(E',M',t)$. The
stationary probability distribution corresponds to an eigenvector
with the largest eigenvalue $1$. The second largest eigenvalue,
$\lambda$, determines the largest relaxation time in the system,
$\tau=\frac{-1}{N \ln \lambda}$. The division by $N$ is needed in
order for $\tau$ to be expressed in units of numbers of
Monte-Carlo steps per total number of spins.The  relaxation times
increase as the system size increases as $\tau\sim L^z$ thus being
characterized by a dynamic exponent, $z$.

\begin{table}
\caption{ Sub-dominant eigenvalues of transition matrices for
different system sides, $L$, and Glauber dynamics. The second
column lists values for the matrix $W(\vec \sigma, \vec \sigma')$
taken from Ref. \cite{Nightingale96,Nightingale00}. The third and
fourth columns are our results for the matrices $T(M;M')$ and
$T(E,M;E'M')$ respectively. \label{tab:1}}
\begin{ruledtabular}
\begin{tabular}{llll}
$L$&$\lambda^W$Ref \cite{Nightingale96,Nightingale00}& $\lambda^{T(M;M')}$ & $\lambda^{T(E,M;E',M')}$ \\
\hline \hline

3 & 0.997409385126011\footnotemark[1] &  0.9973901755 &
0.99740630184576\footnotemark[1]\\
    &   &   & 0.9974063007 \\ \hline
4 &   0.999245567376453\footnotemark[1] & 0.9992429803 &
0.99924409354918\footnotemark[1]\\
    &   &   & 0.9992441209 \\ \hline
5 &  0.999708953624452\footnotemark[1] & 0.9997066202 &
0.99970673172786\footnotemark[1]\\
    &   &   & 0.9997067351 \\ \hline
6&0.9998657194&0.9998635780&0.9998637800\\ \hline
7&0.9999299708&0.9999281870&0.9999284453\\ \hline
8&0.9999600854&0.9999586566&0.9999589090\\ \hline
9&0.9999756630&0.9999744986&\\ \hline
10&0.9999843577&0.9999834244&\\ \hline
11&0.9999895056&0.9999887396&\\ \hline
12&0.9999927107&0.9999921039&\\ \hline
13&0.9999947840&0.9999942741&\\ \hline
14&0.9999961736&0.9999957520&\\ \hline
15&0.9999971315&0.9999967823&\\ \hline
16&0.9999978080&0.9999975119&\\ \hline
17&0.9999982987&0.9999980505&\\ \hline
18&0.9999986606&0.9999984474&\\ \hline
19&0.9999989315&0.9999987550&\\ \hline
20&0.9999991370&0.9999989750&\\ \hline
21&0.9999992955&0.9999991723&\\ \hline 30&&0.9999998016&\\

\end{tabular}
\end{ruledtabular}
\footnotetext[1]{Exact}
\end{table}

We studied the critical behavior of the projected dynamics at the
critical point of the square lattice Ising model, $\beta J=
\frac{1}{2} \ln (1+\sqrt 2)$ for Glauber and Metropolis {\sl et
al.} acceptance probabilities. Our eigenvalue results for the
matrix $T(M,M')$, $\lambda^{T(M,M')}$ and for the matrix
$T(E,M;E'M')$, $\lambda^{T(E,M;E'M')}$ for the Glauber dynamics
can be seen in Table \ref{tab:1} together with the results for the
full state space dynamics, $\lambda^{W}$, obtained from
\cite{Nightingale96,Nightingale00} for the Glauber dynamics using
a variational method. For small systems $L=3,4,5$ we have computed
$\lambda^{T(E,M;E'M')}$ from an exact enumeration of all the
system states and the results are in close agreement with the ones
obtained from the Monte-Carlo estimation of $T_\infty(E,M;E',M')$.
The eigenvalues, for a given system side, $L$  are close to each
other and are observed to obey the inequality $\lambda^{W}>
\lambda^{T(E,M;E'M')}>\lambda^{T(M;M')}$.

\begin{figure}
\includegraphics{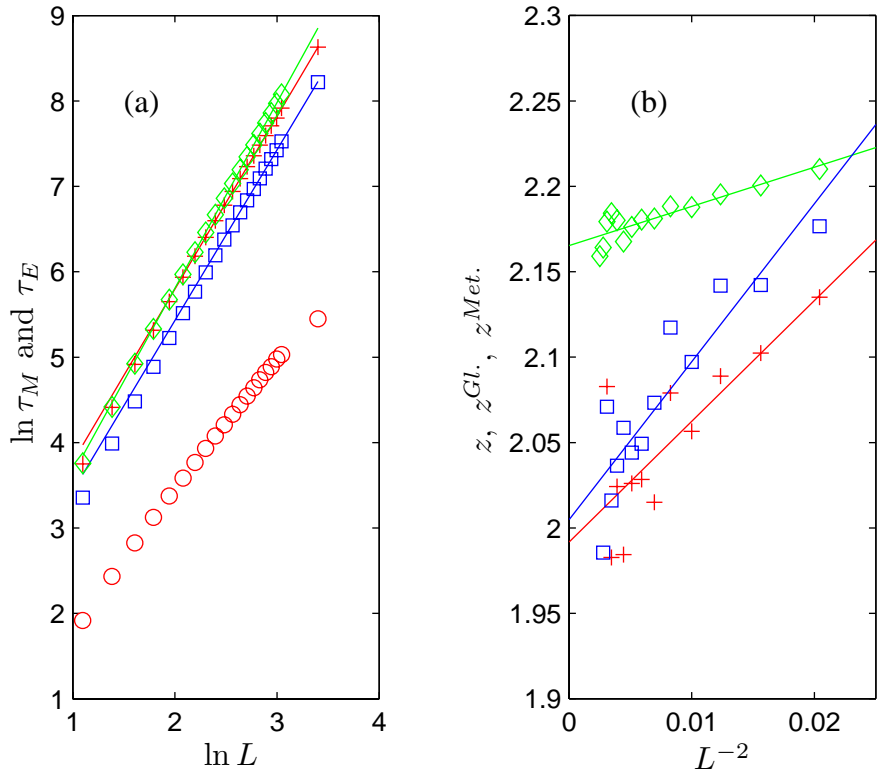}
\caption{\label{fig:1}(a) Values of $\ln \tau_M$ for Glauber
dynamics obtained from the results reported in Ref.
\cite{Nightingale00} ($\diamond$) and values obtained from
$T(M;M')$ (+) as a function of $\ln L$. We also plot $\ln \tau_M$
for the Metropolis {\sl et al.} dynamics ($\Box$) and $\tau_E$ for
the Glauber dynamics obtained from the matrix $T(E;E')$ ($\circ$).
The fitted straight lines were obtained neglecting data for $L<15$
and have slopes $z^{Gl.}=2.02$, $z^{Met.}=2.00$ and $z=2.18$. (b)
Estimations  of the dynamic critical exponent from the local
slopes of the graph in (a) as a function of $L^{-2}$. The symbols
are as in (a). The extrapolated exponents are $z^{Gl.}=1.991$,
$z^{Met.}=2.00$ and $z=2.165$. }
\end{figure}

In Fig. \ref{fig:1} (a) we plot the logarithm of the relaxation
time $\ln \tau_M$, as a function of $\ln L$, for the Glauber
dynamics,  and  the Metropolis {\sl et al.} dynamics, obtained
from the sub-dominant eigenvalue of $T(M;M')$, together with the
full state space results of \cite{Nightingale96,Nightingale00} and
also $\tau_E$ for the Glauber dynamics, obtained from $T(E;E')$.
The fitted straight lines were obtained neglecting data for $L<15$
and have slopes $z^{Gl.}=2.02$, $z^{Met.}=2.00$ and $z=2.18$. To
estimate a reliable value of the exponent $z$ a careful analysis
is needed taking into account corrections to scaling. In
\cite{Nightingale00}the authors report their best estimate
$z=2.1660(10)$  excluding the Domany conjecture $z=2$ with a
logarithmic factor $\tau \sim L^2 (1+b \ln L)$\cite{Domany}.
Further analysis\cite{Arjunwadkara}, by other authors, of the data
of Nightingale and Blothe was not able to categorically exclude
the validity of the Domany conjecture.  The logarithmic dependence
of $\tau_E$ on system size, in accordance with previously reported
results\cite{Wang99}, can be seen in Fig. \ref{fig:1} (a). We have
not tried to make a detailed analysis of our results in order to
have precise estimates of the dynamic exponent from the
magnetization projected dynamics. However, in Fig. \ref{fig:1} (b)
we show the local slope $z=\frac{\ln (\tau(L+1)/\tau(L))}{\ln
((L+1)/L)}$ as a function of $L^{-2}$ which is the first order
finite size correction to the leading behavior considered in
\cite{Nightingale96,Nightingale00}, $\tau\sim L^z(1+ b L^{-2})$.
The results of the extrapolation to the infinite system size limit
shown in  figure Fig \ref{fig:1} (b) are $z^{Gl.}=1.99$,
$z^{Met.}=2.00$ and $z=2.165$. The results for $z^{Gl.}$ and
$z^{Met.}$ seem  to be consistent with $z^{Gl.}\sim z^{Met.}\sim
2$.

\begin{figure}
\includegraphics{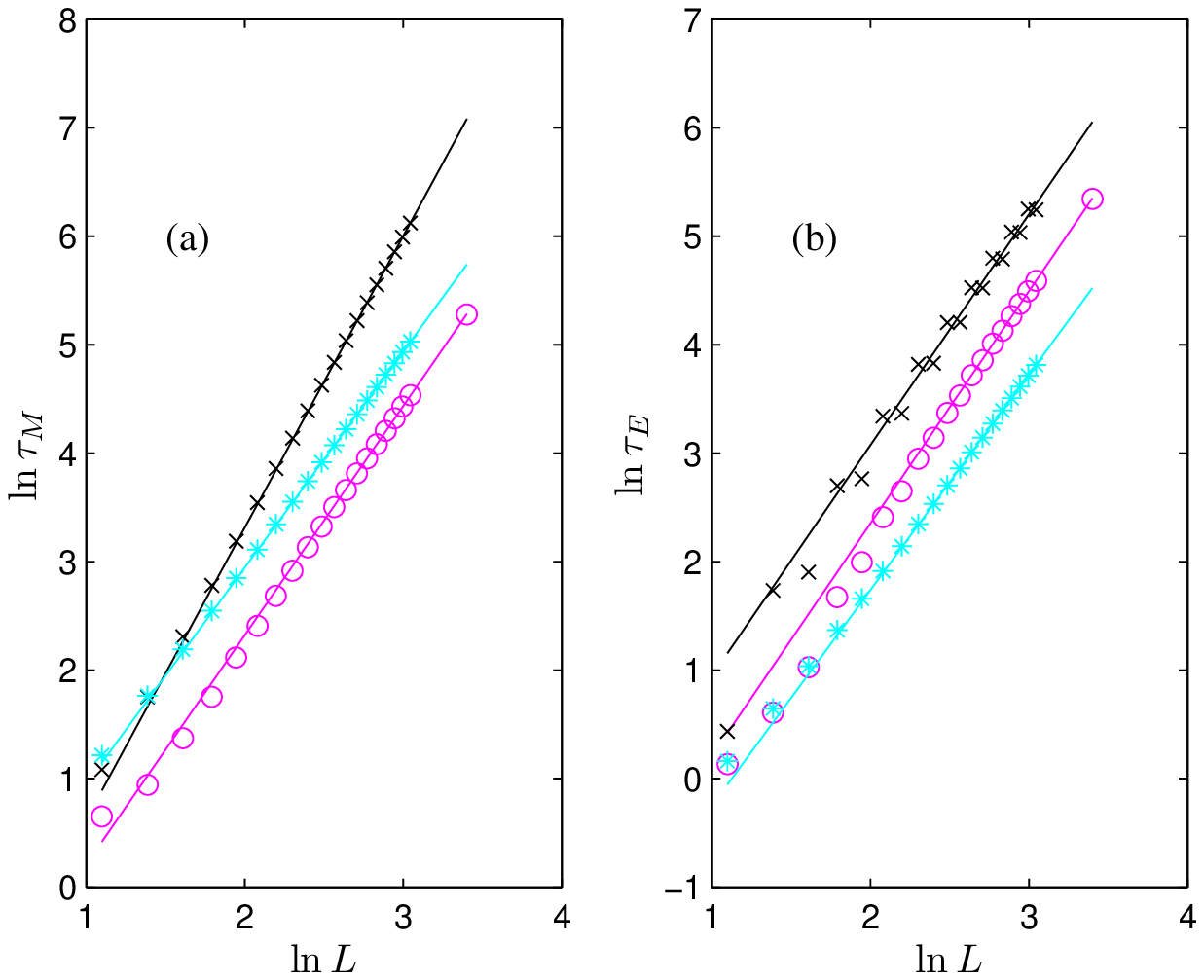}
\caption{\label{fig:2}Relaxation times obtained for the Metropolis
{\sl et al.} dynamics in the flat energy-magnetization ensemble
($\circ$), flat magnetization ensemble ($*$) and flat energy
ensemble ($\times$). In (a) we plot $\tau_M$ obtained from
$T(M;M')$ and the fitted straight lines have slopes
$z^M_{E,M}=2.11$, $z^M_E=2.69$ and $z^M_{M}=1.99$.  In (b) we plot
$\tau_E$ obtained from $T(E;E')$ and the fitted straight lines
have slopes $z^E_{E,M}=2.14$, $z^E_E=2.13$ and $z^E_{M}=1.99$.
Both in (a) and (b) data for $L<15$ were neglected in the fits. }
\end{figure}

In Fig. \ref{fig:2}(a) we plot the relaxation times obtained from
$T(M;M')$ for the Metropolis {\sl et al.} dynamics in the flat
energy-magnetization ensembles, flat energy ensemble and flat
magnetization ensemble. The fitted straight lines have slopes
given by  $z^M_{E,M}=2.11$, $z^M_E=2.69$ and $z^M_{M}=1.99$,
respectively. In Fig. \ref{fig:2}(b) we plot the relaxation times
obtained from $T(E;E')$ for the Metropolis {\sl et al.} dynamics
in the same ensembles. The slopes of the fitted straight lines are
$z^E_{E,M}=2.14$, $z^E_E=2.13$ and $z^E_{M}=1.99$, respectively.
We neglected the data for $L< 15$ in all the fits shown in Figs.
\ref{fig:2} (a) and (b).

\begin{figure}
\includegraphics{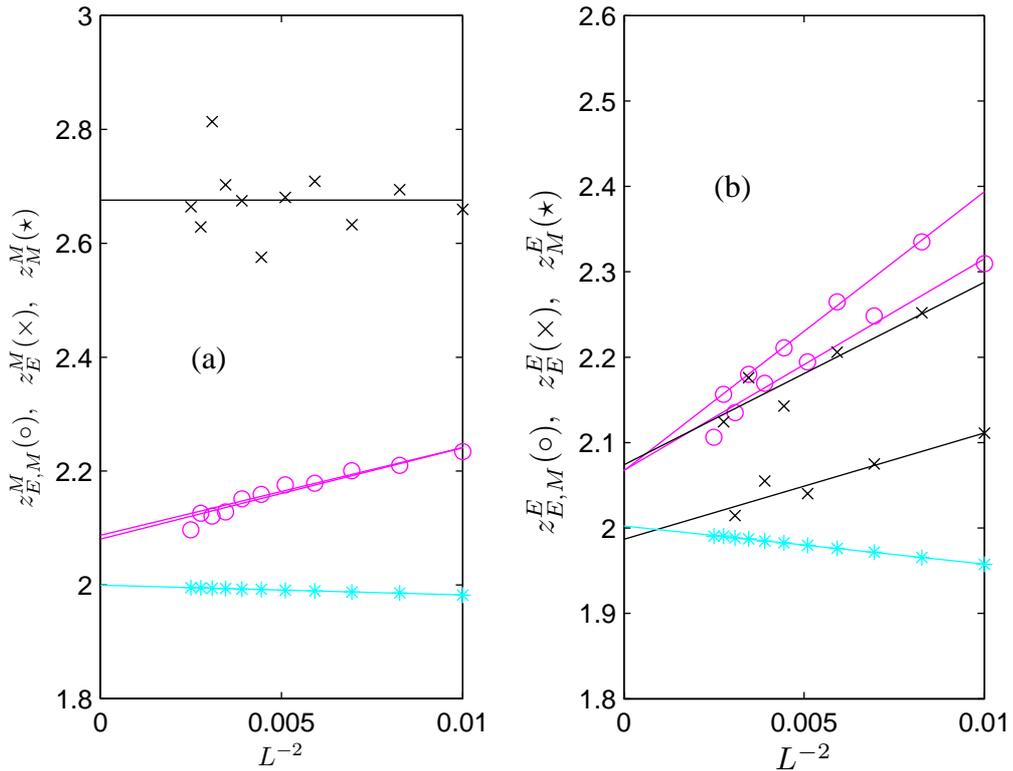}
\caption{\label{fig:21}(a) Dynamic exponent estimates $z_{E,M}^M
(\circ), \ \ z_E^M (\times), \ \ z_M^M(*)$ from local slopes of
the plots shown in Fig. \ref{fig:2} (a) as a function of $L^{-2}$.
For the estimates for the energy-magnetization flat multicanonical
ensemble we made separate estimations for even and odd $L$ system
sides. The infinite system size extrapolations give,
$z_{E,M}^M=2.08$, $z_E^M=2.68$ and $z_M^M=2.00$. (b) $z_{E,M}^E
(\circ), \ \ z_E^E (\times), \ \ z_M^E(*)$ from local slopes of
the plots shown in Fig. \ref{fig:2} (b) as a function of $L^{-2}$.
For the estimates for the flat $(E,M)$ ensemble and for the flat
energy ensemble we made separate estimations for even and odd $L$
system sides. The infinite system size extrapolations give,
$z_{E,M}^E=2.07$, $z_E^E=2.07 $ and $1.99$ for odd and even L,
respectively, and $z_M^E=2.00$.}

\end{figure}

In Fig. \ref{fig:21}(a) we show the estimation of the dynamic
exponent from the local slopes of the relaxation time plots in the
magnetization projected space (Fig. \ref{fig:2}(a)) as a function
of $L^{-2}$. The estimates for $z_E^M$ (flat energy histogram) in
Fig. \ref{fig:21}(a) seem to have converged for the system sizes
studied. The average value for system sides $L\ge 15$ is
$z^M_E=2.68(2)$. This result is  compatible with the
available\cite{guerra2004} result $z=2.80(13)$ obtained by a Monte
Carlo estimate of the convergence time of the time-dependent
energy histogram to the stationary flat distribution of the
energy. The estimates for  $z_M^M$ (flat magnetization histogram)
plotted in the same figure show a dependence with system size
approaching a value close to 2 for infinite system sides. The
infinite size extrapolation for $z_{E,M}^M$ (flat energy and
magnetization histogram) gives a value slightly larger than 2,
$z^M_{E,M}=2.08$. For this multicanonical dynamics the results
show an even-odd effect and it is important to do separate
estimates for even and odd system sides. In Fig. \ref{fig:21}(b)
we also show the estimation of the dynamic exponent from the local
slopes of the plots (Fig. \ref{fig:2}(b)) for the energy projected
dynamics. The infinite size extrapolation for $z_{E,M}^E$ for even
and odd system side are very close to each other and given by,
$z_{E,M}^E=2.07$. The extrapolations for $ z_E^E $ for odd system
side and even system side give $z_E^E=2.07 $ and
$1.99$,respectively. The difference between these two estimates
may be a sign of the presence of corrections to scaling not
properly accounted by our analysis. The estimates for $z_M^E$ show
a size dependence that is compatible with a value close to 2.

The full state transition rate matrix $W$, in the flat
magnetization ensemble  is lumpable with respect to the
classification of the states according to their magnetization.
Consequently, the result $z^M_{M}=2.00$ does not suffer from the
approximation inherent to the projection procedure. A sufficient
and necessary condition for lumpability\cite{Kemeny76}, is that
the total probability to go from a state belonging to a given
magnetization class to another class with different magnetization
is the same for every state in the starting class. For each state
in the starting class with magnetization M there are $n_\pm$
states in the final class $M\pm 2$ where $n_\pm$ is the number of
up/down spins in the initial configuration. The probability to
move to each of these final states in the final class has a
constant value that depends only on the initial $M$ and on the
final $M\pm 2$. All the states in the starting class have the same
number of up spins and down spins so the probability to move to
$M\pm 2$ is the same for every state in the starting class. The
matrix $T(M;M')$ is a tridiagonal symmetric matrix with matrix
elements given by, $T(M+2;M)=T(M;M+2)=\frac{n_+ + 1}{N}$ for
$M<0$, and $T(M+2;M)=T(M;M+2)=\frac{n_-}{N}$ for $M\ge 0$.

\section{Magnetization tunnelling times}
\label{sectionV}

As a measure of performance for multicanonical methods the average
tunnelling times were introduced\cite{dayal2004}. These tunnelling
times measure the time required to sample the whole phase space
and scale with system size differently than the relaxation time.
It was shown that it is important to distinguish between
tunnelling from ground-sate to maximum energy, the up direction,
and from the high energy to the ground-state, the down
direction\cite{costa2005}.

All the tunnelling times reported by us are calculated for the
projected dynamics associated with $T(M;M')$. We calculate the
average time, $\tau_t$ for the system to go from magnetization
$M=-N$ to $M=N$.  We also consider two other average times: The
time $\tau_u$ for the system to go from $M=-N$ to zero
magnetization,  and the time $\tau_d$ for the system to go either
to $M=+N$ or $M=-N$ when it starts from $M=0$. This definition of
$\tau_u$ and $\tau_d$ apply only to systems with even $L$ (and
$N$) such that $M=0$ is an accessible value of the magnetization.

The tunnelling times above defined obey the relation
$\tau_u+\tau_d=\tau_t/2$ that follows from the following simple
argument: For the system to go from $M=-N$ to $M=N$ it has to
reach $M=0$ at some point. It will do so for the first time using
an average time $\tau_u$. Then with probability $1/2$ it will
reach for the first time $M=N$ and the tunnelling time would be
$\tau_u+\tau_d$ or it will return to $M=-N$ and it will reach
later $M=N$ taking a time $\tau_t$. Consequently the tunnelling
times obey the relation
$\frac{1}{2}(\tau_u+\tau_d)+\frac{1}{2}\tau_t=\tau_t$. This
argument uses the fact that the matrix $T(M;M')$ has the symmetry
property $T(M\pm2;M)=T(-M\mp 2;-M)$ and so the walk along positive
values of the magnetization has the same statistical properties of
the walk along negative values of the magnetization.

The time to go from $M=-N$ to $M=N$ can be easily computed taking
advantage of the fact that $T(M;M')$ is non zero only when
$M=M'\pm 2$ and $M'=M$. If we do not allow  transitions from $M=N$
to $M=N-2$ the  $M=N$ becomes an absorbing site for every walk
along the magnetization axis meaning that it will end there upon a
first visit. Defining $h(M)$ as the average time  spent at
magnetization value $M$ \cite{Novotny01}, we can write:
\begin{equation}
h(M-2) T(M;M-2)-h(M) T(M-2;M)=1 \label{eq:20}
\end{equation}
which means that the difference between the average number of
jumps in the positive direction ($M-2 \rightarrow M$) and the
average number of jumps in the negative direction ($M \rightarrow
M-2$) should be equal to one since the system will eventually
reach $M=N$ by moving one time in excess in the positive direction
through the bond connecting the sites $M-2$ and $M$. At $M=N$
there are no jumps in the negative direction and so $h(N-2) T(N;
N-2)=1$. It is then simple to calculate $h(M)$ and the average
tunnelling time for the system to go from $M=-N$ to $M=N$ is given
by,$ \tau_t=\sum_{M=-N}^{M=N-2} h(M)$.

The time $\tau_u$ to reach for the first time $M=0$ starting from
$M=-N$ is obtained using the recursion (\ref{eq:20}) together with
the equation $h(-2) T(0; -2)=1$ to get $\tau_u=\sum_{M=-N}^{M=-2}
h(M)$. Finally, the average time required to start from $M=0$ and
reach for the first time either $M=-N$ or $M=N$, $\tau_d$ is
obtained from the recursion
\begin{equation}
h(M) T(M-2;M)- h(M-2) T(M;M-2)=1 \label{eq:21}
\end{equation}
with a modified rate $T(-2;0)$ equal to $T(-2;0)+T(2;0)$ and
$h(-N+2) T(-N;-N+2)=1$. The average time $\tau_d$ is then given by
$\tau_d=\sum_{M=-N+2}^{M=0} h(M)$. The average tunnelling times
obtained by this method could also have been obtained from the
calculation of the probability of first visit to the absorbing
site that can be computed from the eigenstates and eigenvectors of
the associated absorbing Markov chain matrix ( see
\cite{costa2005}).

The tunnelling times are characterized by dynamic
exponents\cite{dayal2004}, $\tau_t\sim L^{d+z_t}$, $\tau_u\sim
L^{d+z_u}$, $\tau_d\sim L^{d+z_d}$. The relation between these
tunnelling times imply that $z_t$ is equal to the biggest of the
two exponents, $z_u$ and $z_d$, $z_t=\max(z_d,z_u)$. Note that the
tunnelling times reported by us are measured in units of lattice
sweeps and not in units of single site updates.

\begin{figure}
\includegraphics{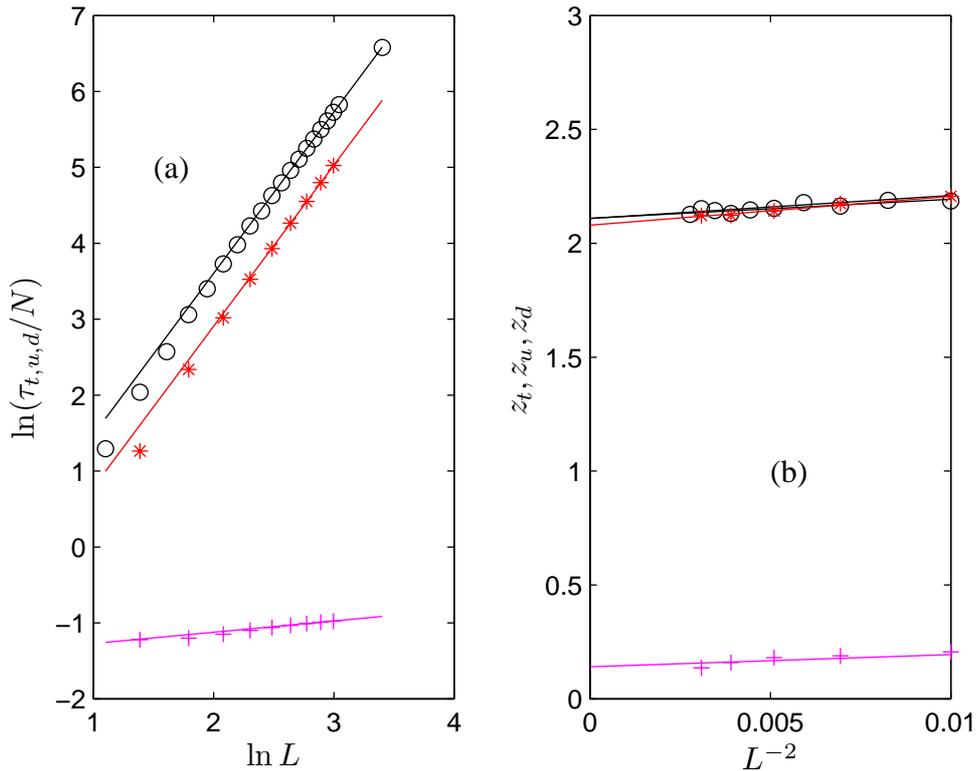}
\caption{\label{fig:3}(a) Tunnelling times, $\tau_t$ ($\circ$),
$\tau_u$(+) and $\tau_d$ (*) as a function of system size $L$ for
the Metropolis {\sl et al.} dynamics in a flat
magnetization-energy histogram ensemble obtained from the matrix
$T(M;M')$. The fitted straight lines were obtained neglecting data
for $L<15$ and have slopes, $z_t=2.12$, $z_u=0.15$ and $z_d=2.12$,
respectively.  In (b) we plot the corresponding local slopes as a
function of $L^{-2}$. Even and odd system sides were treated
separately. The infinite system extrapolation gives, $z_t=2.11$,
$z_u=0.14$ and $z_d=2.08$. }
\end{figure}

In Fig. \ref{fig:3}(a) we show the size dependence of the
tunnelling times, $\tau_t$, $\tau_u$ and $\tau_d$ for the
Metropolis {\sl et al.} dynamics in a flat magnetization-energy
histogram ensemble obtained from the matrix $T(M;M')$. We see that
$\tau_t \sim \tau_d \gg \tau_u$. The scaling exponent of $\tau_u$
obtained from the plot is $z_u=0.15$ which predicts a scaling
$\tau_u \sim L^{2.15}$ close to the exponent of the relaxation
time $z^M_{E,M}=2.11$ reported in the previous section. This
behavior is similar to the one found in \cite{costa2005} where
$\tau_u$ (in the energy space) was found to scale like the
relaxation time of the system. For the other two exponents we have
obtained $z_t\sim z_d \sim2.12$. Note that for a random walk in
the magnetization axis a value for these exponents equal to 0 is
expected. In Fig. \ref{fig:3}(b) we show the local slopes for the
plots in Fig. \ref{fig:3}(a) as a function of $L^{-2}$. The
estimates for $z_t$ and $z_d$ seem to follow a straight line
predicting an infinite system value  $2.11$ and $2.08$,
respectively. The infinite size extrapolation for $z_u$ is $0.14$.

\begin{figure}
\includegraphics{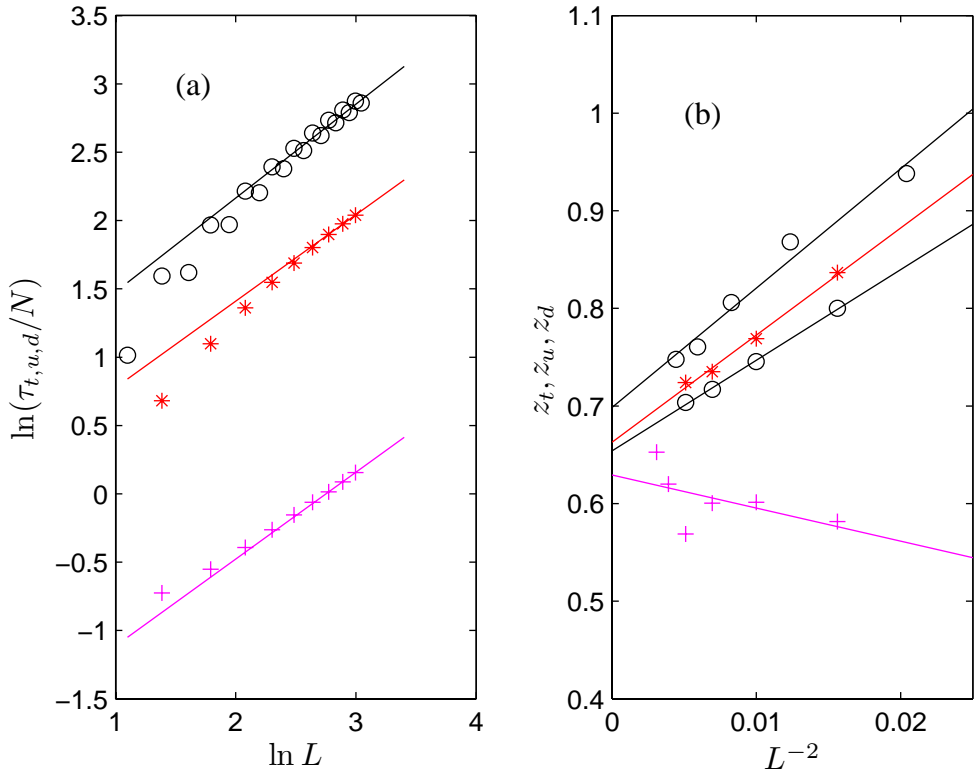}
\caption{\label{fig:4}(a) Tunnelling times, $\tau_t$ ($\circ$),
$\tau_u$(+) and $\tau_d$ (*) as a function of system size $L$  for
the Metropolis {\sl et al.} dynamics in a flat energy histogram
ensemble obtained from the matrix $T(M;M')$. The fitted straight
lines were obtained neglecting data for $L<15$ and have slopes,
$z_t=0.69$, $z_u=0.64$ and $z_d=0.63$, respectively. In (b) we
plot the corresponding local slopes as a function of $L^{-2}$. For
the $z_t$ estimates even and odd system sides were treated
separately. The infinite size extrapolations are, $z_t=0.70$ and
0.65 for odd and even system sides, respectively, $z_u=0.63$ and
$z_d=0.66$.}
\end{figure}

In Fig. \ref{fig:4}(a) we show the size dependence of the
tunnelling times, $\tau_t$, $\tau_u$ and $\tau_d$ for the
Metropolis {\sl et al.} dynamics in a flat energy histogram
ensemble obtained from the matrix $T(M;M')$. The slopes of the
fitted straight lines give $z_t=0.69$, $z_u=0.64$ and $z_d=0.63$.
The result for $z_t$  can be compared with the value $0.78$
reported in Ref. \cite{vianalopes2006} and the value $0.743(7)$
reported in Ref. \cite{dayal2004} by measuring average times for
energy excursions.  An exponent $z_u=0.6$, also obtained from
Monte-Carlo estimates of energy tunnelling times was previously
reported\cite{tesejoao} in very good agreement with our result. In
Fig. \ref{fig:4}(b) we make infinite size extrapolations giving,
$z_t=0.70$ and 0.65 for odd and even system sides, respectively,
$z_u=0.63$ and $z_d=0.66$.

\begin{figure}
\includegraphics{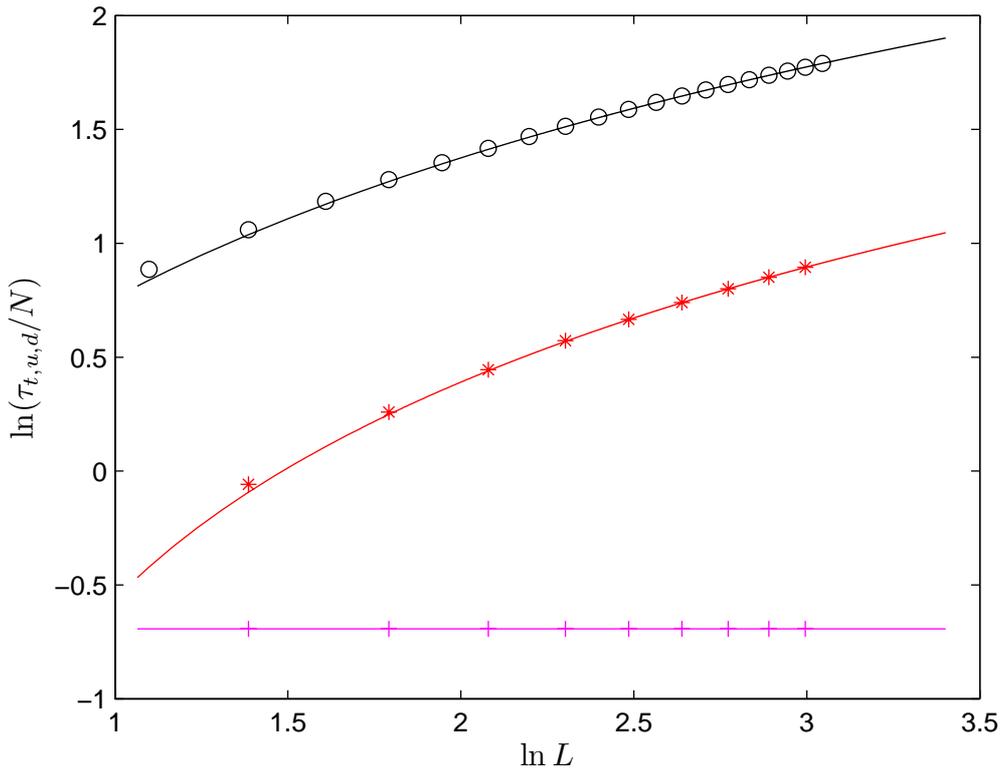}
\caption{\label{fig:5}(a) Tunnelling times, $\tau_t$ ($\circ$),
$\tau_u$(+) and $\tau_d$ (*) as a function of system size $L$  for
the Metropolis {\sl et al.} dynamics in a flat magnetization
histogram ensemble obtained from the matrix $T(M;M')$. The lines
are the analytical asymptotic results given in the text.}
\end{figure}

Finally, we consider the Metropolis {\sl et al.} dynamics for the
flat magnetization histogram ensemble. For this case it is
possible to compute analytically the tunnelling times from the
recursion relations given above, Eqs. (\ref{eq:20},\ref{eq:21})
and the knowledge of the matrix $T(M;M')$. The analytical results
are, $\tau_u=N/2$, $\tau_t=(N+1) H(N/2)$ and
$\tau_d=\frac{1}{2}\bigl ((N+1) H(N/2)-N \bigr)$ where,
$H(n)=\sum_{k=1}^n 1/k$ is the Harmonic number. Using the known
asymptotic result, for large $n$, $H(n)\sim (\ln n + \gamma)$
where $\gamma=0.5772156649...$ is the Euler constant we have
asymptotic expressions for the tunnelling times that predict,
$\tau_t/N\sim \tau_d /N \sim \ln N$ and the tunnelling  exponents
are, $z_t=z_d=z_u=0$. In Fig. \ref{fig:5}(a) we compare the
numerical results for the tunnelling times, $\tau_t$, $\tau_u$ and
$\tau_d$  with the analytical results. Note that, because of the
logarithmic dependence of $\tau_t$ and $\tau_u$ the estimates for
the exponents $z_t$ and $z_d$ that we could obtain for the slopes
of the data shown in Fig. \ref{fig:5}(a) give effective values
around $0.35$ that would slowly approach zero only if larger
systems were considered.

From the three multicanonical ensembles studied we see that the
flat magnetization ensemble is the one with smaller tunnelling
exponents and relaxation time exponent. Recently, it was shown
that it is possible to optimize the ensemble in multicanonical
simulations such that the tunnelling exponent, $z_t$ is also
reduced to zero\cite{trebst2004,vianalopes2006}.

\section{Concluding Remarks}
\label{sectionVI} We have shown that projected dynamics in the
magnetization space is a reasonably good approximation to the full
state space single spin flip dynamics studied in this work:
canonical ensemble Glauber and Metropolis {\sl et al.} dynamics
and three multicanonical ensemble dynamics with flat
energy-magnetization, flat energy and flat magnetization
histograms. The energy projected dynamics is generally a worse
approximation being not able to preserve the power-law size
increase of the relaxation time for canonical ensemble dynamics.
From all the studied dynamics only the flat energy histogram
dynamics show a z exponent clearly larger than 2 and near $2.7$.
For the case of the flat magnetization histogram the projection in
the magnetization space is exact and it is possible to obtain
analytical results for the tunnelling times predicting a zero
value for the  exponents, $z_t, z_d$ and $z_u$. The tunnelling
exponents, $z_t$ (and $z_d$) for the energy and magnetization flat
histogram ensemble are much bigger, $z_t= z_d\sim 2$ and larger
than the exponent $z_u \sim 0$. For the flat energy histogram
dynamics these three exponents are not very different and the
estimates fall between the values, $z_u \sim 0.63$ and $z_t\sim
0.70$ for odd system sides. These results were obtained from the
tunnelling properties of the projected dynamics in the
magnetization space that were found to be in rough agreement with
ones obtained by independent methods for excursions in the energy
space for the flat energy multicanonical ensemble.

Finally, the results  show that the evaluation of the relative
performance of single-spin flip dynamics in Ising like models can
be done very efficiently by studying the projected dynamics in the
magnetization space: the approximation gives reasonably accurate
dynamic exponents;  any, arbitrary, single-spin flip dynamics can
be studied from Monte-Carlo estimations of $T_{\infty}(E,M;E',M')$
for several system sizes in the energy-magnetization space and the
large dimensional reduction achieved by the projection in the
magnetization space allows the application of matrix
diagonalization techniques for bigger system sizes.

Furthermore, the application of projection methods to cluster
dynamics in Ising models and also to other models projected along
their slowest mode may be of considerable interest.

\begin{acknowledgments}
We acknowledge financial support by the MEC (Spain) and FEDER (EU) through
projects FIS2006-09966 and  FIS2004-953. A L C Ferreira thanks the portuguese Funda\c c\~ao para a
Ci\^encia e Tecnologia (FCT) for finantial support.
\end{acknowledgments}

\end{document}